# Comparative studies of the scanning tunneling spectra in cuprate and iron-arsenide superconductors

N-C Yeh[1], M L Teague[1], A D Beyer[1], B Shen[2] and H-H Wen[2]

[1] Department of Physics, California Institute of Technology, Pasadena, CA 91125, USA
[2] Department of Physics, Nanjing University, China

E-mail: ncyeh@caltech.edu

**Abstract**. We report scanning tunneling spectroscopic studies of cuprate and iron-arsenic superconductors, including $YBa_2Cu_3O_{7-\delta}$ (Y-123, $T_c$ = 93 K), $Sr_{0.9}La_{0.1}CuO_2$ (La-112, $T_c$ = 43 K), and the "122" compounds $Ba(Fe_{1-x}Co_x)_2As_2$ (Co-122 with $x$ = 0.06, 0.08, 0.12 for $T_c$ = 14, 24, 20 K). In zero field ($H$ = 0), spatially homogeneous coherence peaks at energies $\omega = \pm \Delta_{SC}$ flanked by spectral "shoulders" at $\pm\Delta_{eff}$ are found in hole-type Y-123. In contrast, only a pair of spatially homogeneous peaks is seen in electron-type La-112 at $\pm \Delta_{eff}$. For $H > 0$, pseudogap ($\Delta_{PG}$) features are revealed inside the vortices, with $\Delta_{PG} = [(\Delta_{eff})^2 - (\Delta_{SC})^2]^{1/2} > \Delta_{SC}$ in Y-123 and $\Delta_{PG} < \Delta_{SC}$ in La-112, suggesting that the physical origin of $\Delta_{PG}$ is a competing order coexisting with superconductivity. Additionally, Fourier transformation (FT) of the Y-123 spectra exhibits two types of spectral peaks, one type is associated with $\omega$-dependent quasiparticle interference (QPI) wave-vectors and the other consists of $\omega$-independent wave-vectors due to competing orders and ($\pi,\pi$) magnetic resonances. For the multi-band Co-122 compounds, two-gap superconductivity is found for all doping levels. Magnetic resonant modes that follow the temperature dependence of the superconducting gaps are also identified. These findings, together with the $\omega$- and $x$-dependent QPI spectra, are consistent with a sign-changing $s$-wave pairing symmetry in the Co-122 iron arsenides. Our comparative studies suggest that the commonalities among the cuprate and the ferrous superconductors include the proximity to competing orders, antiferromagnetic (AFM) spin fluctuations and magnetic resonances in the superconducting (SC) state, and the unconventional pairing symmetries with sign-changing order parameters on different parts of the Fermi surface.

## 1. Introduction

Superconductivity (SC) has been one of the most intellectually challenging topics in condensed matter physics in the past century. The discovery of high-temperature superconducting (high-$T_c$) cuprates in 1986 and the recent discovery of iron-based superconductors in 2008 further defy conventional wisdom to avoid oxides and magnetic materials in search of high-$T_c$ superconductors. In fact, the interplay of magnetism and superconductivity has been a common theme among a variety of unconventional superconductors, from strongly correlated cuprates [1,2] to intermediately correlated ferrous superconductors [3,4], and to nearly itinerant heavy-fermion superconductors [5]. For instance, the parent state of cuprate superconductors is an AFM Mott insulator [1], with common AFM spin fluctuations and magnetic resonances within the $d_{x^2-y^2}$-wave SC phase upon doping carriers into the $CuO_2$ planes [6]. Similarly, a variety of $d$-wave heavy-fermion superconductors exhibit a coexisting

AFM phase and a magnetic resonance associated with collective spin-1 excitations [7]. The ferrous superconductors are also in proximity to AFM metals or semiconductors [3,4], and a magnetic resonance has been observed in inelastic neutron scattering studies of the "122" iron arsenides Ba(Fe$_{1-x}$Co$_x$)$_2$As$_2$ and (Ba$_{1-x}$K$_x$)Fe$_2$As$_2$ [8,9]. A universal relationship between the magnetic resonant mode and the SC gap among the aforementioned systems has also been proposed [10].

In this work we report our comparative studies of cuprate and ferrous superconductors by means of scanning tunneling spectroscopy (STS) as a function of magnetic field ($H$), energy ($\omega$), wave-vector ($q$) and doping level ($x$). Our investigation reveals evidence for the coexistence of competing orders in the $d$-wave SC state of both hole- and electron-type cuprates, and the competing energy scale is larger than the SC gap in the former and smaller than the SC gap in the latter. This finding naturally accounts for the presence of pseudogap (PG) only in the hole-type cuprates and suggests competing orders being the physical origin of the PG. Additionally, we observe field-induced spectral enhancement of the ($\pi,\pi$) magnetic resonance in the cuprates. In contrast, spectroscopic studies of the 122 iron-arsenic compounds reveal direct evidence for two SC gaps and magnetic resonant modes in zero fields. These findings together with detailed analysis of the quasiparticle interference (QPI) spectra provide strong support for a sign-changing $s$-wave pairing symmetry in the multi-band iron-arsenic compounds. The physical implications of our comparative studies of the cuprates and ferrous compounds are discussed.

## 2. Experimental

The samples investigated in this work include optimally doped single crystalline YBa$_2$Cu$_3$O$_{7-\delta}$ (Y-123, $T_c$ = 93 K), optimally doped and high-pressure synthesized polycrystalline Sr$_{0.9}$La$_{0.1}$CuO$_2$ (La-112, $T_c$ = 43 K), and single crystalline Ba(Fe$_{1-x}$Co$_x$)$_2$As$_2$ (Co-122 with $x$ = 0.06, 0.08, 0.12 for $T_c$ = 14, 24, 20 K, corresponding to underdoped, optimally doped and overdoped systems). All samples had been characterized by x-ray diffraction (XRD), magnetization and electrical transport studies to ensure single-phase structures, sharp superconducting transitions and high sample quality [11-13].

For scanning tunneling spectroscopic studies, the surface of the cuprates was prepared by chemical etching [14-17], and samples were kept either in high-purity helium gas or under high vacuum at all times. This surface preparation technique for the structurally highly three-dimensional Y-123 and La-112 cuprates has the advantage of terminating the sample top surface at the metallic CuO$_2$ plane according to the XPS studies [14], thus yielding reproducible spectra with atomic resolution achievable in the constant-bias conductance maps [15-17]. In the case of surface preparation for iron arsenides [18], we performed mechanical cleavage of the single crystals in pure argon atmosphere at room temperature, well above the tetragonal-to-orthorhombic structural phase transition to minimize the commonly observed surface reconstruction for samples cleaved at cryogenic temperatures. The cleaved samples were loaded in situ onto the cryogenic probe of our homemade, high-field (up to 7 T) compatible scanning tunneling microscope (STM) in argon. The sealed STM assembly was subsequently evacuated and cooled to 6 K in UHV with a base pressure at ~ 10$^{-10}$ Torr.

At each constant temperature ($T$) and magnetic field ($H$), spatially resolved tunneling conductance ($dI/dV$) vs. energy ($\omega = eV$) spectra for the quasiparticle local density of states (LDOS) maps were obtained by tunneling along the crystalline c-axis under a range of bias voltages at a given location. The typical junction resistance was ~1GΩ. Current ($I$) vs. voltage ($V$) measurements were repeated pixel-by-pixel over an extended area of the sample. To remove slight spatial variations in the tunnel junction resistance, the ($dI/dV$) at each pixel is normalized to its high-energy background [15-18].

## 3. Results

We illustrate in figure 1 representative tunneling conductance spectra of (a) Y-123 for $H$ = 0 (main panel), $H$ = 2 T (upper inset) and $H$ = 6 T (lower inset), (b) La-112 for $H$ = 0 (main panel) and $H$ = 1 T (inset), and (c) an overdoped Co-122 system with $x$ = 0.12 for $H$ = 0 (main panel) and an optimally doped 112 system with $x = 0.08$ for $H = 0$ (inset), all taken at $T = 6$ K. For $H = 0$, spatially homogeneous coherence peaks at energies $\omega = \pm\Delta_{SC}$ ($\sim \pm 21$ meV) flanked by spectral "shoulders" at $\pm\Delta_{eff}$ ($\sim \pm 38$ meV) are found in Y-123. Assuming coexisting competing orders with SC and using the analysis outlined in Section 4 and detailed previously [2,19,20], we find the relationship $(\Delta_{eff})^2 = (\Delta_{SC})^2 + (\Delta_{PG})^2$, where $\Delta_{PG} \sim 32$ meV is the PG associated with a competing order energy. In contrast, only a pair of spatially homogeneous peaks is seen in electron-type La-112 at $\pm\Delta_{eff}$ ($\sim \pm 13$ meV). For $H > 0$, $\Delta_{PG}$ features are revealed inside the vortices, with $\Delta_{PG} \sim 32$ meV $> \Delta_{SC}$ in Y-123 and $\Delta_{PG} \sim 8$ meV $< \Delta_{SC}$ in La-112. The empirical value of $\Delta_{PG}$ obtained from the PG inside the vortex core is in excellent agreement with that derived from theoretical fitting to the zero-field spectra (solid lines) under the assumption that the competing energy gap is $\Delta_{PG}$. Moreover, inside the vortex cores of Y-123, additional features at a small energy scale $\pm\Delta' \sim \pm 10$ meV are revealed, as exemplified in the left inset of figure 1(a). These features become more pronounced with increasing $H$ [15,16].

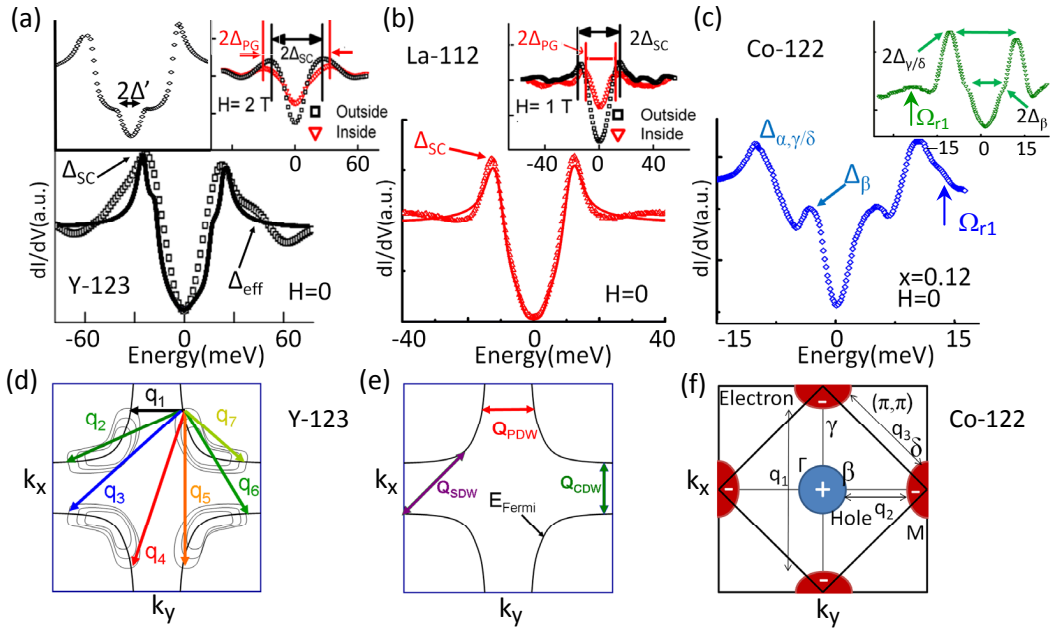

**Figure 1.** (a) – (c) Representative tunnelling spectra of superconductors taken at $T = 6$ K: (a) Y-123 for $H = 0$ (main panel), 2 T (left inset) and 6 T (right inset); (b) La-112 for $H = 0$ (main panel) and 1 T (inset); (c) Co-122 for $x = 0.12$ (main panel) and $x = 0.08$ (inset) with $H = 0$. The solid lines in (a) and (b) are theoretical fitting curves [2,15,16]. (d) – (f) Two-dimensional (2D) Fermi surfaces (FS) of the superconductors: (d) The QPI wave-vectors $\mathbf{q}_i$ due to elastic scattering of quasiparticles in cuprate superconductors [2,15,16]; (e) The low-energy excitation wave-vectors $\mathbf{Q}$ due to CDW, PDW and SDW in the cuprates; (f) The 2D FS of Co-122 in the one-iron unit cell, showing the presence of $\alpha$ and $\beta$ hole pockets at the $\Gamma$-point of the Brillouin zone and electron pockets $\gamma$ and $\delta$ at the M-points. The SC order parameters are opposite in sign for the hole and electron pockets. Possible QPI wave-vectors $\mathbf{q}_1$, $\mathbf{q}_2$ and $\mathbf{q}_3$ connecting different parts of the FS are indicated.

For both underdoped and overdoped Co-122, our surface preparation routinely yielded fragmented surfaces with atomic steps and no apparent surface reconstruction [18]. The resulting LDOS spectra exhibit distinct two SC gaps over $\sim 90\%$ of the sample surface [18], as exemplified in the main panel of figure 1(c) for an overdoped Co-122 with $x = 0.12$, where $\Delta_{\alpha,\gamma/\delta} \sim 10$ meV and $\Delta_\beta \sim 5$ meV. In contrast, applying the same surface preparation procedure for the optimally doped samples resulted in mostly reconstructed surfaces ($\sim 85\%$), and the LDOS spectra revealed predominantly one large gap $\Delta_{\gamma/\delta} \sim 14$ meV, with only weak features of a secondary gap $\Delta_\beta \sim 7$ meV in $\sim 15\%$ of the spectra, as exemplified in the inset of figure 1(c). The mechanism for the surface reconstruction occurring primarily in the optimally doped Co-122 is unclear. We also show in figure 1 (d) – (f) the Fermi surfaces (FS) and the relevant wave-vectors due to quasiparticle interferences (QPI) and the low-energy excitations of competing orders for the cuprate and the ferrous superconductors.

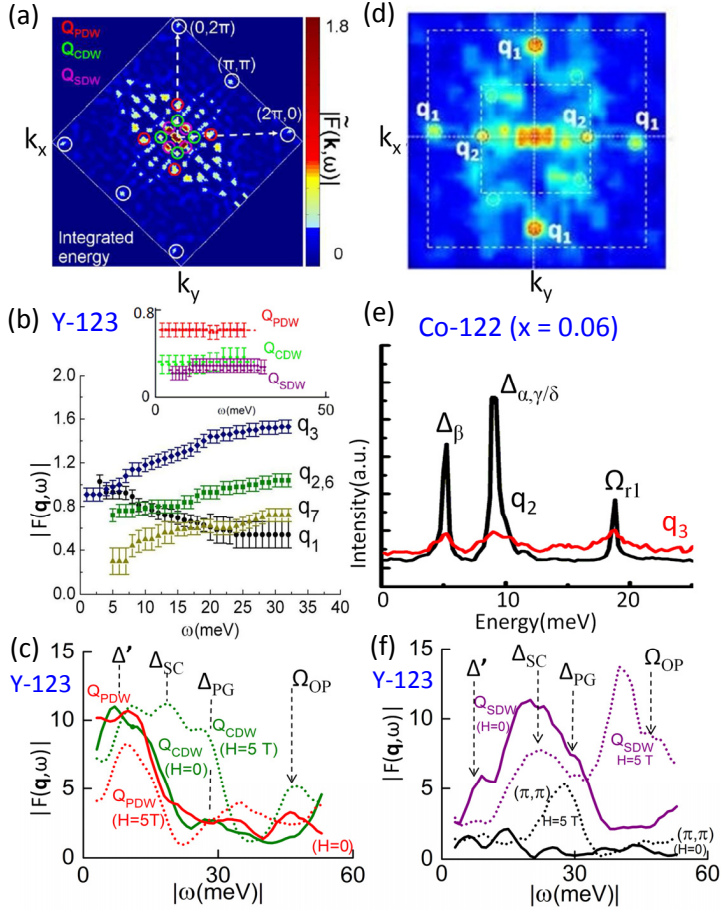

**Figure 2.** FT-LDOS spectroscopic studies of Y-123 and Co-122 at $T = 6$ K: (a) The FT-LDOS intensity $\mathcal{F}(\mathbf{q},\omega)$ (colour) of Y-123 for $H = 5$ T in the 2D reciprocal space and integrated from $\omega = 0$ to 45 meV; (b) The QPI wave-vectors $|\mathbf{q}_i|$ in Y-123 are shown as a function of $\omega$. The inset shows the $\omega$–independent competing order wave vectors $|\mathbf{Q}_{XDW}|$ where X = C, P, and D. (c) The FT-LDOS intensities $\mathcal{F}(\mathbf{q},\omega)$ of Y-123 for $\mathbf{q} = \mathbf{Q}_{CDW}$ and $\mathbf{Q}_{CDW}$ are shown as a function of $\omega$ and for $H = 0$ and $H = 5$ T. (d) $\mathcal{F}(\mathbf{q},\omega)$ of underdoped Co-122 ($x = 0.06$) for $H = 0$ at $\omega = \Delta_{\alpha,\gamma/\delta} \sim 8$ meV is shown in the 2D reciprocal space. Strong intensities at $\mathbf{q} = \mathbf{q}_2$ and $\mathbf{q}_1$ together with an additional nematic order are found [5]. (e) $\mathcal{F}(\mathbf{q},\omega)$-vs.-$\omega$ of underdoped Co-122 ($x = 0.06$) for $H = 0$ at $\mathbf{q} = \mathbf{q}_2$, showing sharp peaks only at $\omega = \Delta_\beta$, $\Delta_{\alpha,\gamma/\delta}$ and $\Omega_{r1}$. (f) The FT-LDOS intensity $\mathcal{F}(\mathbf{Q}_{SDW},\omega)$ and $\mathcal{F}(\mathbf{Q}_{(\pi,\pi)},\omega)$ of Y-123 are shown as a function of $\omega$ and for $H = 0$ and $H = 5$ T.

Next, we perform studies of the Fourier transformation (FT) of the LDOS spectra to deduce wave-vectors associated with the quasiparticle interferences (QPI) and low-energy collective excitations. The FT-LDOS of Y-123 integrated from the Fermi level ($\omega = 0$) to $\omega = 45$ meV is shown in figure 2(a) over the reciprocal space, and the $\omega$-dependence of various wave-vectors (**q**) identified from the high-intensity spots in (a) is illustrated in figure 2(b), with the energy-dependent $|\mathbf{q}|$-values shown in the main panel and the energy-independent $|\mathbf{q}|$-values given in the inset. The energy-dependent $|\mathbf{q}|$-values are in good agreement with theoretically predicted QPI wave-vectors due to elastic scattering of quasiparticles [2,15,16], whereas the $\omega$-independent wave-vectors may be attributed to the momentum of charge-, pair- and spin-density waves (CDW, PDW and SDW) [15,16]. Empirically, $\mathbf{Q}_{CDW} = [\pm(0.28\pm0.02)\pi,0] / [0,\pm(0.28\pm0.02)\pi]$, $\mathbf{Q}_{PDW} = [\pm(0.56\pm0.06)\pi,0] / [0,\pm(0.56\pm0.06)\pi]$, and $\mathbf{Q}_{SDW} = [\pm(0.15\pm0.01)\pi, \pm(0.15\pm0.01)\pi]$. Moreover, $\mathbf{Q}_{(\pi,\pi)} = (\pi,\pi)$ appears for $H > 0$.

The $\omega$ and $H$-dependent evolution of the FT conductance ($\mathcal{F}$) at specific $|\mathbf{Q}|$-values is exemplified in figure 2(c) for $\mathcal{F}(\mathbf{Q}_{CDW})$-vs.-$\omega$ and $\mathcal{F}(\mathbf{Q}_{PDW})$-vs.-$\omega$ taken at $H = 0$ and 5 T, and in figure 2(f) for $\mathcal{F}(\mathbf{Q}_{SDW})$-vs.-$\omega$ and $\mathcal{F}(\mathbf{Q}_{(\pi,\pi)})$-vs.-$\omega$ taken at $H = 0$ and 5 T. We note that $\mathcal{F}(\mathbf{Q}_{PDW})$ and $\mathcal{F}(\mathbf{Q}_{CDW})$ exhibit intense spectral weight at $\omega \sim \Delta'$ for $H = 0$, the former becomes strongly suppressed for $H = 5$ T, suggesting that $\Delta'$ may be associated with the coherence energy of localized Cooper pairs. In contrast, $\mathcal{F}(\mathbf{Q}_{CDW})$ becomes much enhanced over a wide $\omega$ range with increasing $H$, suggesting that suppressing SC by $H$ increases the particle-hole spectral weight of CDW. On the other hand, the $(\pi,\pi)$ magnetic resonance near $\omega = \Delta_{PG}$ becomes strongly enhanced upon increasing $H$. In the case of $\mathcal{F}(\mathbf{Q}_{SDW},\omega)$, the broad peak near $\Delta_{SC}$ becomes suppressed by increasing $H$, and the spectral weight shifted significantly to energies above $\Delta_{PG}$, implying that $\Delta_{PG}$ may be associated with the CDW/SDW particle-hole excitations. Finally, a local spectral maximum near $\omega = 45 \sim 50$ meV for all $\mathcal{F}(\mathbf{Q}_{XDW},\omega)$ curves suggests possible coupling of the Cu-O optical phonon frequency $\Omega_{op}$ to competing orders.

For comparison, the FT-LDOS of underdoped Co-122 in the reciprocal space for the one-iron unit cell is shown in figure 2(d) for $\omega = \Delta_{\alpha,\gamma/\delta}$. Two QPI wave-vectors $\mathbf{q}_1$ and $\mathbf{q}_2$ are identified [18], and the $\omega$-dependence of $\mathcal{F}(\mathbf{q}_2,\omega)$ is shown in figure 2(e). The pronounced peaks of $\mathcal{F}(\mathbf{q}_2,\omega)$ at $\omega = \Delta_\beta$, $\Delta_{\alpha,\gamma/\delta}$ and $\Omega_{r1} \sim (\Delta_\beta + \Delta_{\gamma/\delta})$ [9] support the notion that $\mathbf{q}_2$ is associated with the QPI wave-vector between the electron- and hole-pockets rather than due to Bragg diffraction because the latter would have been $\omega$-independent. The absence of $\mathbf{q}_3$ in figure 2(d) further corroborates the $s_\pm$-wave pairing [18].

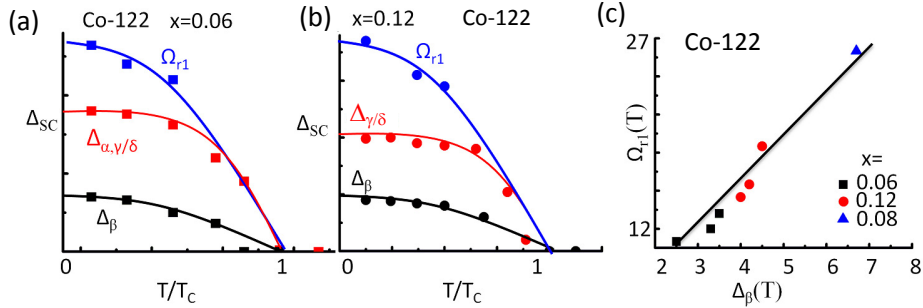

**Figure 3.** Correlation between the SC gaps $\Delta_{\alpha,\gamma/\delta}$, $\Delta_\beta$ and the magnetic resonant mode $\Omega_{r1}$ of the Co-122 superconductors: (a) $T$-dependence of $\Delta_{\alpha,\gamma/\delta}$, $\Delta_\beta$ and $\Omega_{r1}$ for the underdoped sample ($x = 0.06$). (b) $T$-dependence of $\Delta_{\gamma/\delta}$, $\Delta_\beta$ and $\Omega_{r1}$ for the overdoped sample ($x = 0.12$). (c) Correlation of the magnetic resonant mode $\Omega_{r1}(T)$ with the SC gap $\Delta_\beta(T)$ for Co-122 samples with three different doping levels of $x = 0.06, 0.08$ and $0.12$. The slope agrees with the relation $\Omega_{r1} \sim 3\Delta_\beta \sim 1.5\Delta_{\gamma/\delta}$.

## 4. Analysis and discussion

### 4.1. Cuprate superconductors

Our spectral analysis of the cuprates begins with a mean-field Hamiltonian $\mathcal{H} = \mathcal{H}_{SC} + \mathcal{H}_{CO}$ that describes coexisting $d_{x^2-y^2}$-wave SC and a competing order, where the $d_{x^2-y^2}$-wave SC Hamiltonian is:

$$\mathcal{H}_{SC} = \sum_{\mathbf{k},\sigma} \xi_\mathbf{k} c^\dagger_{\mathbf{k},\sigma} c_{\mathbf{k},\sigma} - \sum_\mathbf{k} \Delta_{SC}(\mathbf{k})\left(c^\dagger_{\mathbf{k},\uparrow} c^\dagger_{-\mathbf{k},\downarrow} + c_{-\mathbf{k},\downarrow} c_{\mathbf{k},\uparrow}\right), \quad (1)$$

with $\Delta_{SC}(\mathbf{k}) = \Delta_{SC}(\cos k_x - \cos k_y)/2$, $\mathbf{k}$ denotes the quasiparticle momentum, $\xi_\mathbf{k}$ is the normal-state eigen-energy relative to the Fermi energy, $c^\dagger$ and $c$ are the creation and annihilation operators, and $\sigma = \uparrow, \downarrow$ refers to the spin states. The Hamiltonian $\mathcal{H}_{CO}$ depends on the type of competing orders involved. For CDW, the dominating competing order for hole-type cuprates at $H = 0$ [19-21], $\mathcal{H}_{CDW}$ is given by:

$$\mathcal{H}_{CDW} = -\sum_{\mathbf{k},\sigma} \Delta_{CDW}(\mathbf{k})\left(c^\dagger_{\mathbf{k},\sigma} c_{\mathbf{k}+\mathbf{Q}_{CDW},\sigma} + c^\dagger_{\mathbf{k}+\mathbf{Q}_{CDW},\sigma} c_{\mathbf{k},\sigma}\right), \quad (2)$$

where the wave-vector $\mathbf{Q}_{CDW}$ is parallel to the $CuO_2$ bonding direction $(\pi,0)/(0,\pi)$ as shown in figure 1(e). In the case of SDW, which is the dominating competing order for electron-type cuprates at $H = 0$ [2,21] and may be enhanced in hole-type cuprates for $H > 0$, the Hamiltonian $\mathcal{H}_{SDW}$ becomes:

$$\mathcal{H}_{SDW} = -\sum_{\mathbf{k},\alpha,\beta} \Delta_{SDW}(\mathbf{k})\left(c^\dagger_{\mathbf{k}+\mathbf{Q}_{SDW},\alpha} \sigma^3_{\alpha\beta} c_{\mathbf{k},\beta} + c^\dagger_{\mathbf{k},\alpha} \sigma^3_{\alpha\beta} c_{\mathbf{k}+\mathbf{Q}_{SDW},\beta}\right), \quad (3)$$

where $\mathbf{Q}_{SDW}$ is parallel to $(\pi,\pi)$ as shown in figure 1(e), and $\sigma^3_{\alpha\beta}$ is the $\alpha\beta$ matrix element of the Pauli matrix $\sigma^3$. Assuming coexisting competing orders and SC, we obtain $\Delta_{PG}$ from fitting the zero-field tunneling spectra [19,20] and angle resolved photoemission spectroscopy (ARPES) data [21] and find that $\Delta_{PG}$ agrees with the PG energies revealed inside the vortex cores of Y-123 and La-112 [2,15-17].

*4.2. Ferrous superconductors*
The dominance of the QPI wave-vector $\mathbf{q}_2$ in the FT-LDOS of all Co-122 samples (see figure 2(d)) and the peaks of the FT-LDOS intensity $\mathcal{F}(\mathbf{q}_2,\omega)$ at $\omega = \Delta_\beta$, $\Delta_{\alpha,\gamma/\delta}$ and $\Omega_{r1}$ (see figure 2(e)) are strong evidences for the sign-changing *s*-wave pairing symmetry, because for $H = 0$ only the QPI wave-vectors connecting parts of FS of opposite signs in the SC order parameter can appear in the FT-LDOS. The small and nearly isotropic electron- and hole-pockets associated with the ferrous superconductors further restrict the occurrence of QPI to energies close to the SC gaps and their linear combinations. This is in contrast to the large FS of the cuprates so that the QPI in the cuprates occurs over a much wider range of momentum and energy [2], as exemplified in figure 2(b). We further note that the magnetic resonances satisfy the relationship $\Omega_{r1} \sim (\Delta_\beta + \Delta_{\gamma/\delta}) \sim 1.5\Delta_{\alpha,\gamma/\delta} \sim 3\Delta_\beta$ and $\Omega_{r2} \sim (\Delta_\alpha + \Delta_{\gamma/\delta}) \sim 2\Delta_{\alpha,\gamma/\delta}$ [9]. Therefore, only one magnetic resonance $\Omega_{r1}$ is observed for optimally and overdoped Co-122 due to vanished $\alpha$-pocket, whereas $\Omega_{r2} \sim (16\pm1)$ meV for underdoped Co-122. The correlation between the SC gaps and magnetic resonances is manifested in figure 3(a)-(c) for all doping levels.

## 5. Conclusion
Our comparative spectroscopic studies of the cuprate and ferrous superconductors suggest that the commonalities among these high-$T_c$ superconductors include their proximity to competing orders, the presence of AFM spin fluctuations and magnetic resonances in the SC state, and the unconventional pairing symmetries with sign-changing order parameters on different parts of the Fermi surface.